\documentclass[conference]{IEEEtran}
\IEEEoverridecommandlockouts
\usepackage{cite}
\usepackage{amsmath,amssymb,amsfonts}
\usepackage{algorithm}
\usepackage{algorithmic}
\usepackage{tabularx}
\usepackage{graphicx} 
\usepackage{booktabs} 
\usepackage{textcomp}
\usepackage{xcolor}
\usepackage{hyperref}
\usepackage{subcaption}
\usepackage{xcolor,tcolorbox}
\usepackage{amsmath}
\def\BibTeX{{\rm B\kern-.05em{\sc i\kern-.025em b}\kern-.08em
    T\kern-.1667em\lower.7ex\hbox{E}\kern-.125emX}}
\begin{document}

\title{GitBugs: Bug Reports for Duplicate Detection, Retrieval Augmented Generation, Triage, and More}

\author{\IEEEauthorblockN{Avinash Patil}
\IEEEauthorblockA{
\textit{Hewlett Packard Enterprise}\\
Sunnyvale, USA \\
avinash.patil@hpe.com \\ 
}
\and

\IEEEauthorblockN{Siru Tao}
\IEEEauthorblockA{
    Heinz College of Information Systems and\\
    Public Policy\\
    Carnegie Mellon University\\
    Pittsburgh, USA \\
    sirutao@andrew.cmu.edu
    }

\and
\IEEEauthorblockN{Aryan Jadon}
\IEEEauthorblockA{
\textit{Hewlett Packard Enterprise}\\
Sunnyvale, USA \\
aryan.jadon@hpe.com \\
}
}

\maketitle

\begin{abstract}
Bug reports provide critical insights into software quality, yet existing datasets are often limited in scope, outdated, or lack the structured metadata required for modern machine learning applications. To address these challenges, we present GitBugs—a large-scale, up-to-date dataset comprising over 150,000 bug reports collected from nine actively maintained open-source projects, including Firefox, Cassandra, and VS Code. GitBugs aggregates data from GitHub, Bugzilla, and Jira issue trackers and standardizes key categorical fields to support supervised learning tasks. It also provides predefined train/test splits for duplicate bug detection, along with exploratory analysis notebooks and project-level statistics such as duplicate rates and resolution times. The dataset enables a wide range of software engineering research tasks, including duplicate detection, retrieval-augmented generation, resolution prediction, automated triaging, and temporal analysis. By offering a cross-project, openly licensed, and machine learning–ready resource, GitBugs facilitates reproducible benchmarking and advances research in automated bug report analysis. The dataset and code are publicly available at \url{https://github.com/av9ash/gitbugs/}.\newline
\end{abstract}

\begin{IEEEkeywords}
Bug reports, Software engineering datasets, Duplicate bug detection, Bug triaging, Resolution prediction, Issue tracking, Empirical software engineering, Machine learning, Data curation, Open-source software.
\end{IEEEkeywords}

\section{Introduction}
Bug reports are foundational to software maintenance, reliability, and quality assurance. They capture critical user feedback and developer annotations that help identify defects, propose fixes, and prioritize engineering resources. Despite their importance, publicly available datasets of bug reports often fall short in scale, consistency, and breadth. Many existing resources focus on single projects, lack standardized metadata, or are outdated.

To address these challenges, we introduce \textbf{GitBugs}—a comprehensive, curated dataset containing over 150,000 bug reports spanning nine major open-source projects. We designed this dataset to support the software engineering research community by providing clean, labeled, and diverse bug tracking data. GitBugs enables empirical investigations and development of automated techniques across domains such as duplicate detection, bug triaging, resolution classification, and retrieval augmented generation.

Our key contributions are as follows:
\begin{itemize}
    \item A large-scale dataset of over 150,000 bug reports collected from 9 open-source software projects, covering Github, Jira and Bugzilla ecosystems.
    \item Comprehensive metadata per report, including fields such as \texttt{Summary}, \texttt{Description}, \texttt{Status}, \texttt{Priority}, \texttt{Resolution}, and \texttt{timestamp} information.
    \item Project-wise analytics including duplicate rates, time-to-resolution distributions, and metadata completeness.
    \item Reproducible artifacts: Duplicate Bug reports mappings, exploratory data analysis (EDA) notebooks, model training and, validation scripts.
\end{itemize}

We intend this dataset to serve as a benchmark for research and development in bug report understanding and automation, and a foundation for training large-scale models on realistic, task-specific software engineering data.

\section{Related Work}
The software engineering community has long recognized the value of curated bug report datasets in advancing research areas such as defect prediction, localization, duplicate detection, and reproducibility. Over the years, several influential datasets have been released, each tailored to specific research objectives.

\subsection{Existing Datasets}
One of the earliest large-scale efforts in bug report analysis was by Lamkanfi et al.~\cite{lamkanfi2013eclipse}, who published a dataset comprising over 200{,}000 bug reports from Eclipse and Mozilla. Their work highlighted the evolution of bug discussions, laying a foundation for bug triaging and prioritization research.

To support duplicate bug report detection, the LogPAI team introduced \textit{BugRepo}~\cite{LogPAI_BugHub}, a collection of free-text bug reports from projects such as Firefox and Eclipse JDT. BugRepo has become a benchmark for applying NLP techniques to software engineering tasks by focusing on textual similarity and clustering of duplicates.

Reproducibility has emerged as another core challenge in bug tracking. In response, Wendland et al.~\cite{wendland2021andror2} released \textit{AndroR2}, a dataset of 90 Android bugs accompanied by reproduction steps, scripts, and APKs. This resource has facilitated research on automated bug reporting and empirical reproducibility studies.

Other datasets emphasize code-level analysis. Ferenc et al.~\cite{ferenc2020automatically} introduced \textit{BugHunter}, which links buggy and fixed code segments with a comprehensive set of code metrics, enabling deeper investigation into defect prediction.

Cross-language bug localization has also gained traction. Muvva et al.~\cite{muvva2020bugl} developed \textit{BuGL}, a multilingual dataset containing over 10{,}000 bug reports from C, C++, Java, and Python projects, aimed at evaluating language-agnostic localization models.

Recent large-scale mining efforts have further enriched the landscape. Oliva et al.~\cite{vieira2019reports} curated a longitudinal dataset mapping bug reports to fix commits across 55 Apache projects, providing insights into long-term bug lifecycles. Complementing this, \textit{RegMiner}~\cite{song2022regminer} automatically collects reproducible regression bugs from 66 projects, supporting research in regression detection and automated repair.

Several other widely-used datasets also exist. The PROMISE repository~\cite{boetticher2007promise} provides datasets for defect prediction, while Defects4J~\cite{just_defects4j_2014} offers reproducible Java bugs. However, these are often limited in scope, with fewer projects and less metadata, such as duplicate links or resolution details.

Our proposed dataset, \textbf{GitBugs}, addresses these limitations by aggregating bug reports from diverse platforms (GitHub, Jira, and Bugzilla) across multiple projects. By offering datasets of varying sizes per repository, it supports flexible training setups—ranging from smaller datasets suitable for current cost-constrained large language model applications to larger datasets ideal for training traditional machine learning models at scale.

\subsection{Applications in Research}

Bug report datasets have supported a variety of research efforts, including duplicate detection~\cite{sun2011towards, patil2024comparative}, automated triaging~\cite{anvik2006should}, and severity prediction~\cite{lamkanfi2010predicting}. The growing application of natural language processing (NLP) and deep learning in software engineering~\cite{hoang2020cc2vec} has amplified the need for large and small, high-quality datasets.

\textbf{GitBugs} enables such applications by offering realistic, multi-domain bug data with detailed annotations. It supports reproducible experimentation across traditional machine learning and large language model (LLM)-based techniques.

\section{Dataset Description}

\subsection{Data Sources}

The GitBugs dataset aggregates bug reports from nine well-established open-source software projects: \textit{Cassandra}, \textit{Firefox}, \textit{Hadoop}, \textit{HBase}, \textit{Mozilla Core}, \textit{VS Code}, \textit{Seamonkey}, \textit{Spark}, and \textit{Thunderbird}. These projects span multiple domains including distributed systems, browsers, IDEs, and cloud infrastructure. We sourced the reports from these bug trackers using project-specific APIs and scraping utilities.

\subsection{Data Collection Methodology}

We gathered bug reports using a combination of RESTful API calls (for Jira-based systems like Apache projects) and HTML scraping or CSV archive parsing (for Bugzilla-based systems like Firefox and Thunderbird). We filtered reports to exclude non-bug entries such as feature requests or tasks using available metadata fields such as \texttt{issue type}. 

\subsection{Data Volume and Statistics}

Table~\ref{tab:project-stats} summarizes the dataset at the project level, reporting the total number of bug submissions and the percentage of duplicates. \textit{Mozilla Core} contains the most significant number of reports, exceeding 85,000, followed by \textit{VS Code} and \textit{Firefox}. In contrast, projects like \textit{SeaMonkey} and \textit{Cassandra} have comparatively smaller datasets. Duplicate rates show considerable variation: \textit{VS Code} and \textit{Thunderbird} both exceed 25\%, suggesting room for improvement in triage and reporting workflows.

Additional insights not shown in figures are outlined below:

\begin{itemize}
    \item \textbf{Resolution time exhibits a heavy-tailed distribution:} While some bugs are resolved within days, others remain open for months or years.
    \item \textbf{Bug priority correlates with outcomes:} High-priority issues are more frequently resolved as \textit{Fixed} or \textit{Won’t Fix}, while low-priority ones are often closed as \textit{Duplicate} or left unresolved.
    \item \textbf{Report lengths are highly variable:} The median bug description is under 600 characters, though some exceed several thousand characters due to verbose logs or stack traces.
\end{itemize}

\begin{table}[h]
\centering
\caption{Summary statistics across projects in GitBugs.}
\label{tab:project-stats}
\begin{tabular}{lrrr}
\toprule
\textbf{Project} & \textbf{Total Reports} & \textbf{Duplicates} & \textbf{Duplicate Rate (\%)} \\\midrule
Cassandra      & 4,612   & 300   & 6.5 \\
Firefox        & 28,824  & 6,255 & 21.7 \\
Hadoop         & 2,503   & 128   & 5.1 \\
HBase          & 5,403   & 108   & 2.0 \\
Mozilla Core   & 85,673  & 17,899 & 20.9 \\
VS Code        & 32,829  & 9,272 & 28.2 \\
Seamonkey      & 1,076   & 120   & 11.2 \\
Spark          & 20,275  & 497   & 2.5 \\
Thunderbird    & 15,192  & 4,200 & 27.6 \\\bottomrule
\end{tabular}
\end{table}

\begin{figure}[htbp]
    \centering
    \includegraphics[width=\linewidth]{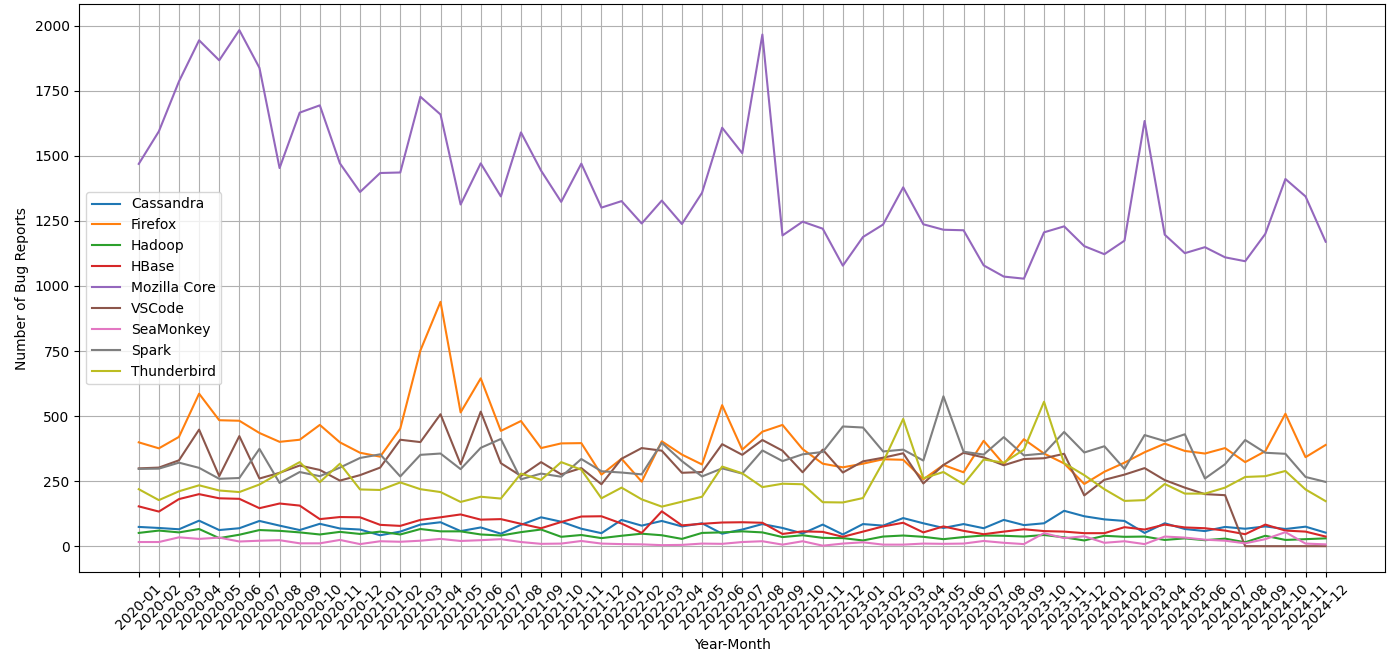}
    \caption{Monthly bug report trends from 2020 to 2024 across multiple projects. Mozilla Core consistently reports the highest volume, while other projects show lower and more variable trends over time.}
    \label{fig:bug_trends}
\end{figure}

Figure~\ref{fig:bug_trends} illustrates monthly bug reporting trends. \textit{Mozilla Core} consistently leads in report volume. \textit{Firefox} shows prominent spikes in mid-2021 and 2024. \textit{VS Code} and \textit{Thunderbird} report at a steady, moderate pace, while \textit{Cassandra} and \textit{SeaMonkey} remain relatively low.

\begin{figure}[htbp]
    \centering
    \includegraphics[width=\linewidth]{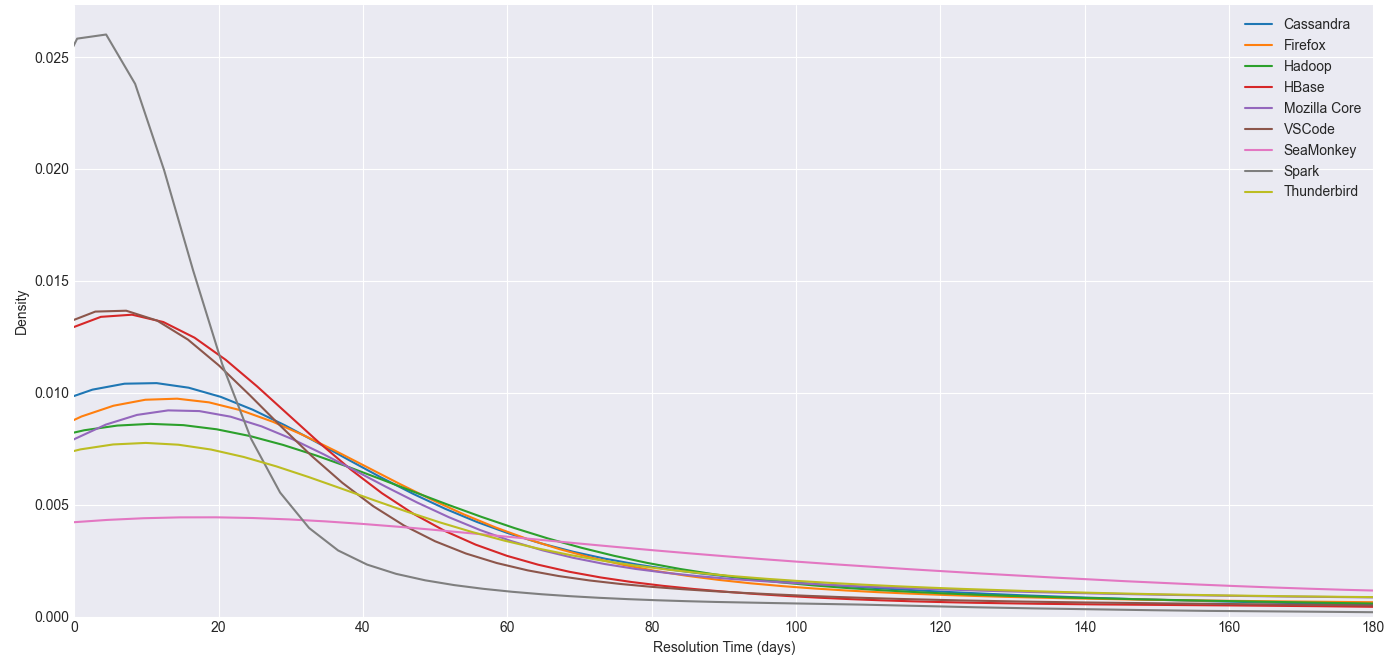}
    \caption{Kernel density estimates of bug resolution times across multiple projects. Spark shows the fastest resolution, while SeaMonkey has the longest tail, indicating slower bug fixes.}
    \label{fig:kde_resolution_time}
\end{figure}

\begin{figure}[htbp]
    \centering
    \includegraphics[width=\linewidth]{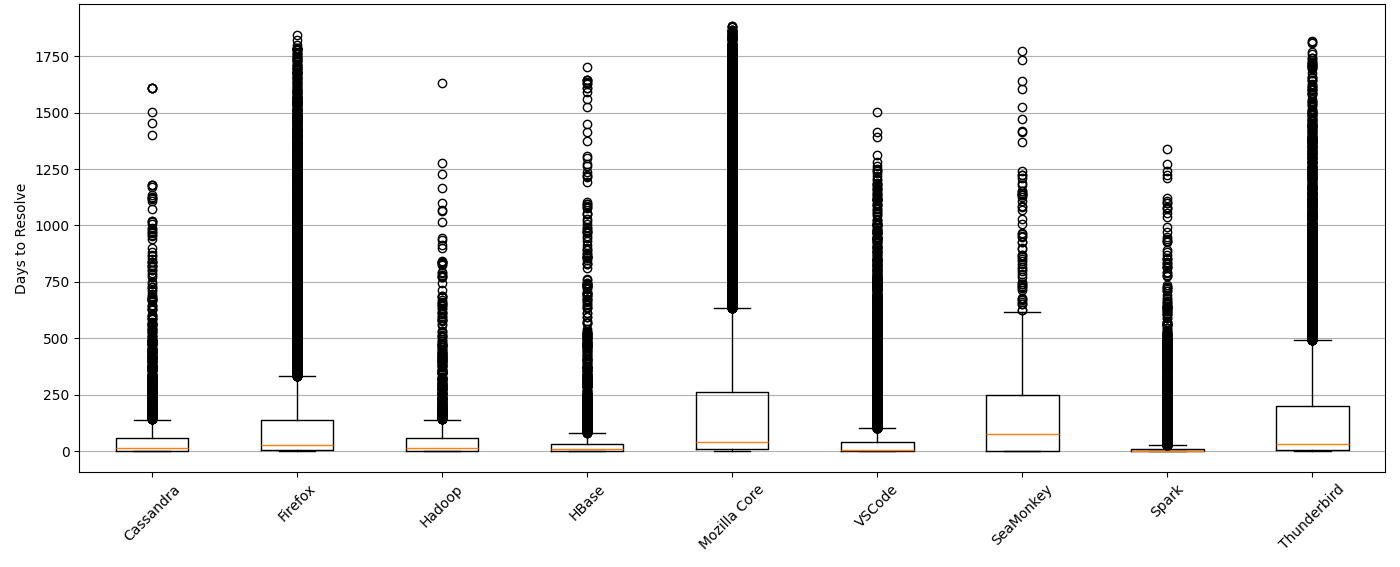}
    \caption{Distribution of bug resolution times across projects using box plots. Most projects exhibit a right-skewed distribution with many outliers; Mozilla Core, SeaMonkey, and Thunderbird show notably longer resolution times.}
    \label{fig:box_resolution_time}
\end{figure}

Figures~\ref{fig:kde_resolution_time} and~\ref{fig:box_resolution_time} compare bug resolution efficiency across projects. \textit{Spark} shows the fastest resolution times, peaking at under 10 days. In contrast, \textit{SeaMonkey} and \textit{Mozilla Core} exhibit longer tails, reflecting extended issue lifetimes. Box plots reveal wide variance in resolution times for projects like \textit{Thunderbird} and \textit{Mozilla Core}, whereas \textit{VS Code} and \textit{Spark} demonstrate more consistent and efficient resolution behavior.

\subsection{Comparison table with existing datasets:}

\begin{table*}[htbp]
\caption{Comparison of Software Bug Report Datasets}
\label{tab:bug_datasets}
\centering
\renewcommand{\arraystretch}{1.2}
\begin{tabular}{@{}p{3cm}p{1cm}p{3.5cm}p{3.5cm}p{4cm}@{}}
\toprule
\textbf{Dataset Name} & \textbf{Year} & \textbf{Source Projects / Domain} & \textbf{Size / Scope} & \textbf{Key Features} \\
\midrule
PROMISE Repository\cite{boetticher2007promise} & 2007 & Various software projects & Empirical software engineering data & Benchmark datasets for software engineering research \\
\midrule
Eclipse \& Mozilla Defect Tracking Dataset\cite{lamkanfi2013eclipse} & 2013 & Eclipse (JDT, CDT, Platform, GEF), Mozilla (Firefox, Thunderbird, Bugzilla, Core) & $\sim$215,000 bug reports & Full bug lifecycle with update history; XML format \\
\midrule
Defects4J\cite{just_defects4j_2014} & 2014 & Java projects & Real faults in Java programs & Enables controlled testing studies \\
\midrule
GitHub Issues Dataset (8M)\cite{github_issues8m} & 2017 & GitHub & 8 million issues & Large-scale dataset for issue tracking and analysis \\
\midrule
Bughub\cite{LogPAI_BugHub} & 2018 & GitHub & Free-text bug reports & Designed for duplicate issue identification \\
\midrule
BugRepo\cite{bugrepo_dataset} & 2018 & Various software projects & Dataset for duplicate bug report detection & Supports research on duplicate bug report detection \\
\midrule
Apache Bug-Fix Commits Dataset\cite{vieira2019reports} & 2019 & 55 Apache projects & 10 years of bug-fix commits & Links bug reports to commits; useful for bug-fix analysis \\
\midrule
BugHunter\cite{ferenc2020automatically} & 2020 & Various open-source Java projects & Code elements (files, classes, methods) with bug information & Automatically constructed bug dataset with code metrics; supports bug prediction research\\
\midrule
Bugl\cite{muvva2020bugl} & 2020 & Cross-language (Java, Python, etc.) & Cross-language bug reports & Facilitates bug localization across languages \\
\midrule
AndroR2\cite{wendland2021andror2} & 2021 & Android apps & Manually reproduced bug reports & Supports reproducibility studies \\
\midrule
Eclipse Bugzilla Dataset\cite{eclipse_dataset} & 2022 & Eclipse & Bug reports from Eclipse's Bugzilla & Structured dataset for bug analysis \\
\midrule
Public Jira Dataset\cite{jira_dataset} & 2022 & Various projects using Jira & Publicly available Jira issues & Facilitates research on issue tracking systems \\
\midrule
RegMiner\cite{song2022regminer} & 2022 & Various code repositories & Replicable regression bug dataset & Focuses on regression bugs \\
\midrule
GitBugs (This Paper) & 2025 & Bugzilla, JIRA, Github & 150K+ Bug Reports with  mapped duplicates& Duplicate Issue Identification, Reterival Augmented Generation, Prediction etc. \\

\bottomrule
\end{tabular}
\end{table*}

\section{Potential Applications}

The GitBugs dataset supports a wide range of research and industrial applications in software engineering, particularly in bug report analysis, natural language processing, and machine learning. Table \ref{tab:bug-survey} presents a brief survey of existing studies on bug report analysis.

\begin{table}[t]
\caption{Survey of traditional bug management tasks and emerging LLM-enabled expansions.}
\centering
\footnotesize
\begin{tabular}{p{4.5cm} p{3cm}}
\toprule
\textbf{Applications} & \textbf{Representative Studies} \\
\midrule

\multicolumn{2}{l}{\textbf{Traditional Tasks}} \\
\midrule

Duplicate Detection 
& \cite{zhang2023cupid, jahan2023towards, mukherjee2025understanding, gotharsson2024exploring, wu2023intelligent, laney2025automated, meng2024combining, jiang2023does, gotharsson2023case, chauhan2023denature, ghadhab2024impact, zheng2024duplicate} \\

Severity Prediction 
& \cite{ali2024bert, zhou2025bug, acharya2024graph, bibyan2024bug, wei2023improving, wu2024crowdsourced, mian2023software, mashhadi2023method, sarawan2023machine, aburakhia2024machine, chowdhury2024method} \\

Component Assignment 
& \cite{samir2023improving, li2024automatic, borg2024adopting, wang2025fixer, xu2023method} \\

Triage Automation 
& \cite{adhikari2025leveraging, zhang2025btal, yadav2023comparison, yerramreddy2023empirical, dipongkor2023comparative, kumar2024ensemble} \\

\midrule
\multicolumn{2}{l}{\textbf{LLM-Enabled Expansions}} \\
\midrule

Report Rewriting \& Clarification 
& \cite{acharya2025can, bo2024chatbr} \\

Missing Information Inference 
& \cite{chen2025empirical, plein2023can} \\

Automated Reproduction Step Extraction 
& \cite{kang2024evaluating, kang2023large, cheng2025agentic, wang2024feedback, wang2025aegis, yin2025bugrepro} \\

Conversational Debugging 
& \cite{chen2025empirical, song2023burt, bajpai2024let} \\

\bottomrule
\end{tabular}
\label{tab:bug-survey}
\end{table}

\subsection{Research Opportunities}

GitBugs supports several high-impact research directions, including:

\begin{itemize}
    \item \textbf{Duplicate bug detection:} The availability of labeled duplicates enables training and evaluation of IR-based, graph-based, and neural models for identifying semantically similar bug reports.
    \item \textbf{Bug triaging:} Supervised models can be developed to predict suitable developers, components, or priorities for new bug reports based on historical assignments and metadata.
    \item \textbf{Resolution and severity prediction:} The structured fields \texttt{Resolution}, \texttt{Priority}, and \texttt{Status} provide valuable ground truth for classification tasks involving severity estimation and outcome forecasting.
    \item \textbf{Time-to-fix modeling:} With timestamp data for creation and resolution dates, regression models can be trained to predict the expected resolution time of a bug.
    \item \textbf{Temporal and linguistic evolution:} Longitudinal analysis of reporting patterns, language use, and resolution behaviors across projects and time can yield insights into software project dynamics and process maturity.
\end{itemize}

\subsection{Industry Use Cases}

Beyond academic settings, the dataset can be leveraged by software organizations for:

\begin{itemize}
    \item \textbf{Tool benchmarking:} Internal bug analysis tools (e.g., triage recommenders or deduplication engines) can be validated using standardized, labeled data from GitBugs.
    \item \textbf{LLM fine-tuning:} The bug summaries and descriptions offer a rich corpus for pretraining or instruction tuning large language models for software maintenance tasks.
    \item \textbf{Custom QA pipelines:} Real-world bug metadata and distributions help design and evaluate risk scoring, prioritization, and anomaly detection models tailored to in-house development workflows.
    \item \textbf{Training and education:} GitBugs can be a hands-on resource in software engineering courses, enabling students to explore realistic bug data and apply analysis techniques.
\end{itemize}

\subsection{Case Study: Trend Analysis and Prediction with GitBugs}
We conducted a case study on Apache Cassandra bug reports to demonstrate the utility of the GitBugs dataset, reserving the final six months of data for testing. The study evaluated various machine learning (ML) models across key software engineering tasks: bug volume forecasting, severity classification, time-to-fix prediction, and trend detection.

\vspace{1em}
\textbf{Prediction Tasks.}

\begin{figure}[htbp]
    \centering
    \includegraphics[width=\linewidth]{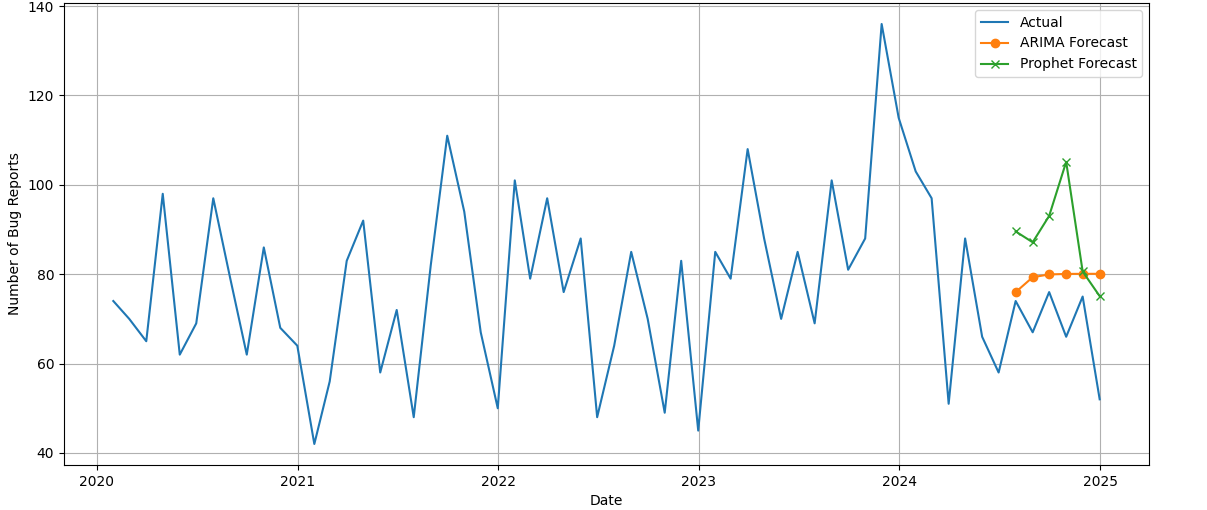}
    \caption{Monthly bug report forecasts using ARIMA and Prophet models. Actual data (blue) is shown alongside ARIMA (orange) and Prophet (green) forecasts for 2024–2025.}
    \label{fig:forecast_comparison}
\end{figure}

\vspace{0.5em}
\textit{Bug Volume Forecasting.} We forecasted future bug report volumes using ARIMA and Prophet models trained on monthly historical data, evaluating the last six months reports. As shown in Figure~\ref{fig:forecast_comparison}, ARIMA produced a stable forecast, while Prophet reflected a sharper recent increase. Based on Mean Absolute Error (MAE), ARIMA outperformed Prophet with scores of \textbf{10.92} and \textbf{20.10}, respectively, underscoring a trade-off between stability and sensitivity to recent trends.

\begin{figure}[htbp]
    \centering
    \includegraphics[width=\linewidth]{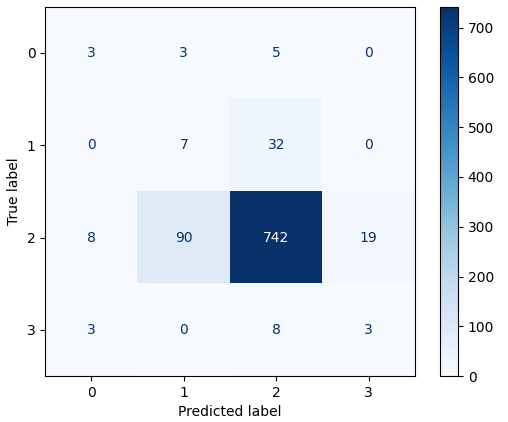}
    \caption{Confusion matrix for bug severity classification on the Cassandra dataset. Most samples are correctly classified as class 2, with moderate confusion between adjacent classes.}
    \label{fig:confusion_matrix}
\end{figure}

\vspace{0.5em}
\textit{Bug Priority Classification.} We trained a multi—class classifier to predict High, Low, Normal, and Urgent bug priorities. As shown in Figure~\ref{fig:confusion_matrix}, it achieved strong performance on the dominant \textit{Normal} class (F1-score: 0.90), while performance on minority classes was poor. Despite an overall accuracy of \textbf{82\%}, the \textbf{macro-averaged F1-score} was only \textbf{0.35}, indicating significant class imbalance and the need for mitigation strategies such as re-sampling or cost-sensitive learning.

\begin{figure}[htbp]
    \centering
    \includegraphics[width=0.85\linewidth]{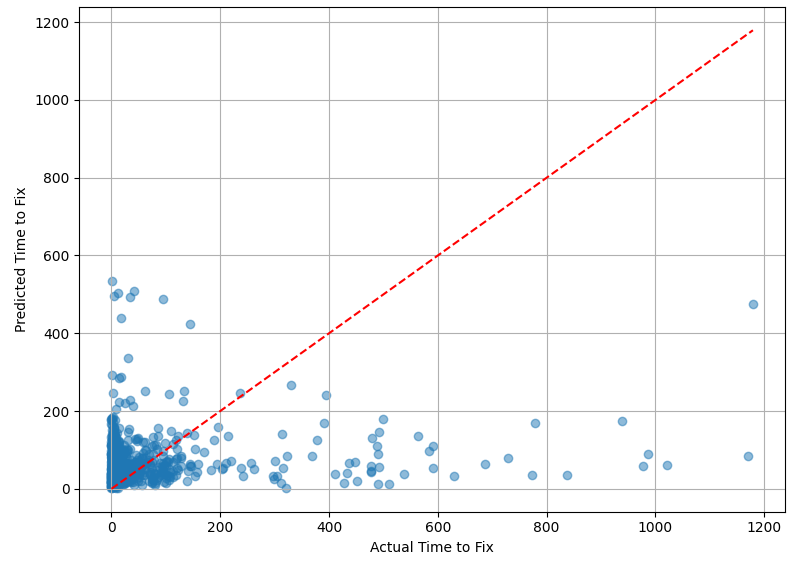}
    \caption{Scatter plot of actual vs. predicted bug fix times. The red dashed line represents the ideal prediction (y = x). Most predictions underestimate the actual time to fix, indicating model bias toward shorter durations.}
    \label{fig:time_to_fix_scatter}
\end{figure}

\vspace{0.5em}
\textit{Time-to-Fix Prediction.} We trained a regression model to predict the time-to-fix for bug reports. As illustrated in Figure~\ref{fig:time_to_fix_scatter}, the predicted values often diverge from the actual values, particularly for reports with longer resolution times. The model yielded a Mean Absolute Error (MAE) of \textbf{86.06}, a Root Mean Squared Error (RMSE) of \textbf{158.44}, and an R\textsuperscript{2} score of \textbf{–0.09}, reflecting poor predictive performance. The negative R\textsuperscript{2} score indicates that the model underperforms even a naive predictor that always returns the mean. These results highlight the need for more informative features or alternative modeling approaches. The use of MAE as a performance metric is consistent with prior work \cite{jadon2024comprehensive}.

\vspace{1em}
\textbf{Trend Analysis.}

\vspace{0.5em}
\textit{Keyword Trends.} We tracked domain-specific terms such as \texttt{repair}, \texttt{auth}, and \texttt{timeout}. Spikes in terms like \texttt{repair} and \texttt{read} correlated with maintenance-focused development phases, revealing shifts in team priorities.

\begin{figure}[htbp]
    \centering
    \includegraphics[width=\linewidth]{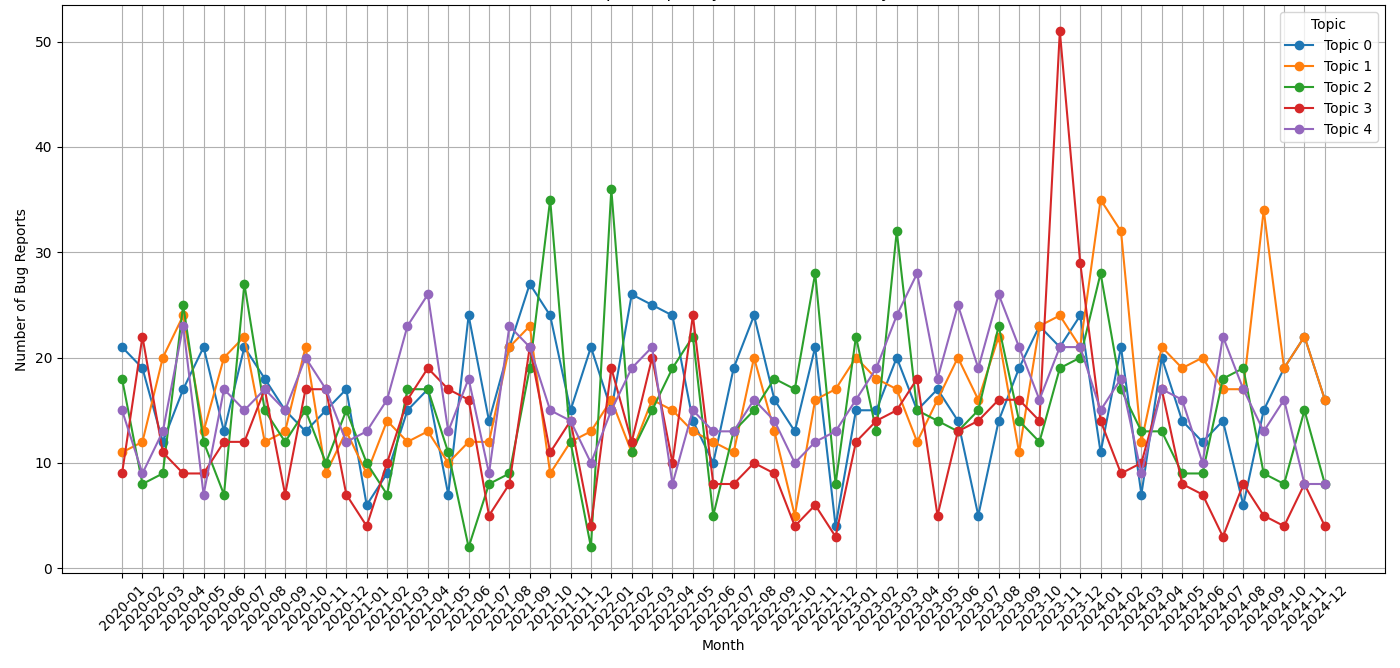}
    \caption{Monthly distribution of bug reports across LDA-derived topics from 2020 to 2024. Each topic shows fluctuating trends, with occasional spikes indicating bursts of activity in specific thematic areas.}
    \label{fig:lda_topics}
\end{figure}

\vspace{0.5em}
\textit{Topic Modeling.} Applying Latent Dirichlet Allocation (LDA) uncovered five dominant topics across the dataset: indexing, test automation, configuration cleanup, CVE/security, and flaky test failures. We illustrated this temporal evolution in Figure~\ref{fig:lda_topics}.

\begin{figure}[htbp]
    \centering
    \includegraphics[width=\linewidth]{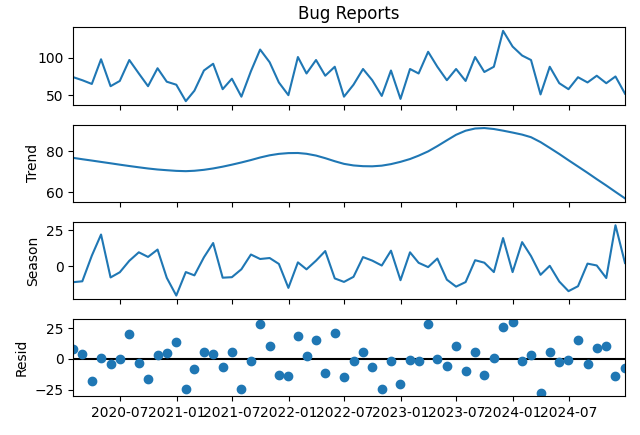}
    \caption{STL decomposition of monthly bug reports: trend, seasonality, and residuals.}
    \label{fig:stl_decomposition}
\end{figure}

\vspace{0.5em}
\textit{Seasonal Decomposition.} We applied STL decomposition to isolate monthly bug volume's seasonal and trend components. The seasonal pattern aligned with quarterly release cycles, while the trend revealed a long-term decline in bug reports—suggesting increased system maturity, as shown in Figure~\ref{fig:stl_decomposition}.

\vspace{0.5em}
\textit{Duplicate Detection.} To evaluate the dataset’s support for information retrieval tasks, we conducted a duplicate bug detection experiment using cosine similarity over bug summaries encoded with Sentence-BERT embeddings. For each of the 300 randomly selected query bugs, We retrieve the top 10 most similar candidates based on cosine similarity scores.

\begin{figure}[htbp]
    \centering
    \includegraphics[width=\linewidth]{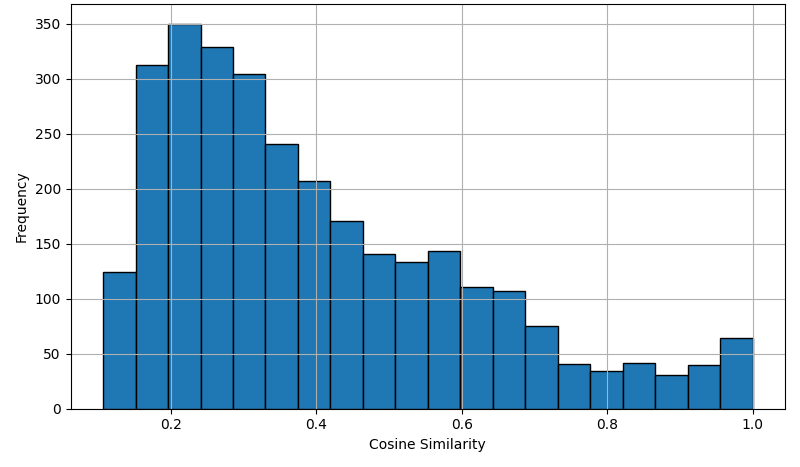}
    \caption{Histogram of cosine similarity scores between bug report embeddings. Most similarities fall below 0.5, indicating substantial semantic variation across reports, with a small cluster of highly similar cases near 1.0.}
    \label{fig:cosine_dist}
\end{figure}

As illustrated in Figure~\ref{fig:cosine_dist}, most top-10 similarity scores fell below 0.5, indicating that many duplicates remain linguistically subtle. Despite this challenge, the approach achieved a Recall@10 of \textbf{0.61}, suggesting that semantic embedding techniques can retrieve a majority of known duplicates with high precision in the candidate set.

\vspace{2em}
\textbf{Retrieval-Augmented Generation.}

\vspace{0.5em}
\textit{Bug Explanation via Retrieval.} To demonstrate GitBugs’ support for natural language workflows, we implemented a Retrieval-Augmented Generation (RAG) pipeline. Given a new bug report, the system retrieves semantically similar historical reports and conditions a language model to produce enriched explanations.

\vspace{0.5em}
\textbf{Input Bug Report:}
\begin{tcolorbox}[colback=gray!5!white, colframe=gray!75!black, boxrule=0.3pt, arc=2pt, left=2pt, right=2pt, top=2pt, bottom=2pt]
\small \texttt{When I press login on the settings page, nothing happens in Firefox Nightly.}
\end{tcolorbox}

\textbf{Retrieved Examples:}
\vspace{-0.5em}
\begin{itemize}[\leftmargin=1.5em]
    \item \texttt{Login button in about:preferences does not trigger any action when clicked.}
    \item \texttt{Firefox login screen fails silently due to a missing DOM event listener on macOS.}
    \item \texttt{Click events not captured on some buttons after upgrading to Nightly 120.0.}
\end{itemize}

\textbf{Generated Explanation:}
\begin{tcolorbox}[colback=blue!1!white, colframe=blue!20!black, boxrule=0.3pt, arc=2pt, left=2pt, right=2pt, top=2pt, bottom=2pt]
\small
\textit{This issue likely stems from a missing or broken event listener tied to the login button. In similar cases, the root cause was a regression in the event handling system introduced in a recent update. Suggested next step: check the associated click handler in the latest Nightly build and verify DOM readiness before event binding.}
\end{tcolorbox}

\noindent This example highlights how GitBugs enables language models to reason with contextual evidence, improving interpretability and alignment with prior developer behavior. Unlike zero-shot generation, RAG leverages historical grounding to support triage, summarization, and root cause hypothesis in a transparent manner.

\vspace{1em}
\textbf{Insights.} This case study highlights GitBugs’ versatility in supporting various analytical tasks. It enables both predictive modeling and exploratory analysis of real-world development patterns. The findings also underscore several challenges, including class imbalance, high variance in fix durations, and limitations of textual signals—providing directions for future research on robust, domain-aware ML methods in software engineering.







\section{Conclusion and Future Work}

\subsection{Summary}

GitBugs represents a scalable, diverse, and richly annotated corpus of real-world software bug reports. By aggregating data from nine widely used open-source projects and unifying their structure, the dataset provides a valuable benchmark for empirical studies and machine learning research in software engineering.

With comprehensive fields such as \texttt{Summary}, \texttt{Description}, \texttt{Status}, \texttt{Priority}, and \texttt{Resolution}, the dataset supports a variety of research tasks including duplicate detection, triaging, severity prediction, and resolution time estimation. We believe GitBugs can facilitate reproducible experimentation and foster software analytics and innovation in tool development.





\bibliographystyle{IEEEtran}
\bibliography{ref}

\end{document}